\documentclass[fleqn,twoside]{article}
\usepackage[headings]{espcrc2}

\readRCS
$Id: espcrc2.tex,v 1.2 2004/02/24 11:22:11 spepping Exp $
\usepackage{graphicx}
\usepackage[figuresright]{rotating}

\newcommand{\AmS}{{\protect\the\textfont2
  A\kern-.1667em\lower.5ex\hbox{M}\kern-.125emS}}

\newcommand{\beq}{\begin{equation}}
\newcommand{\eeq}{\end{equation}}
\newcommand{\bea}{\begin{eqnarray}}
\newcommand{\eea}{\end{eqnarray}}
\newcommand{\beas}{\begin{eqnarray*}}
\newcommand{\eeas}{\end{eqnarray*}}
\newcommand{\ra}{\rightarrow}
\newcommand{\epm}{e^+e^-}
\title{Current status of {\tt carlomat}, a program 
for automatic computation of lowest order cross sections}

\author{K. Ko\l odziej\address{Institute of Physics, University of Silesia\\
          ul. Uniwersytecka 4, PL-40007 Katowice, Poland}%
        \thanks{Supported by the Polish Ministry of Scientific Research 
         and Information Technology as a research grant No. N N519 404034 
         in years 2008--2010 and by European Community's Marie-Curie Research 
         Training Network under contracts MRTN-CT-2006-035482 (FLAVIAnet) 
         and MRTN-CT-2006-035505 (HEPTOOLS).}}
        
\runtitle{Current status of {\tt carlomat}, a program 
for automatic computation of lowest order cross sections}
\runauthor{K. Ko\l odziej}

\begin{document}

\begin{abstract}
The current status of {\tt carlomat}, a program for automatic 
computation of the lowest order cross sections of multiparticle reactions
is described, the results of comparisons with other multipurpose 
Monte Carlo programs are shown and some new results on 
$\epm \ra b \bar b b \bar b u\bar d d \bar u$ are presented.

\vspace{1pc}
\end{abstract}

\maketitle

\section{MOTIVATION}

Many interesting aspects of the Standard Model (SM) and models beyond it
can be studied through investigation of reactions involving a few
heavy particles at a time. Owing to the high energy and luminosity 
such reactions will be copiously observed at particle
colliders such as the Large Hadron Collider (LHC), or the International Linear 
Collider (ILC) \cite{ILC}. As the heavy particles are usually unstable,
they almost immediately decay leading to reactions with several
particles in the final state. Already in the lowest order of 
SM matrix elements of such multiparticle reactions receive 
contributions typically from many thousands of the Feynman diagrams,
most of which constitute background to the ``signal diagrams'' representing
the interesting subprocesses of production and decay of those heavy particles.
Because of the large number of Feynman diagrams involved, reliable SM 
predictions for such reactions can be obtained only through a fully 
automated calculational process. 

\begin{figure*}[htb]
\vspace{100pt}
\includegraphics{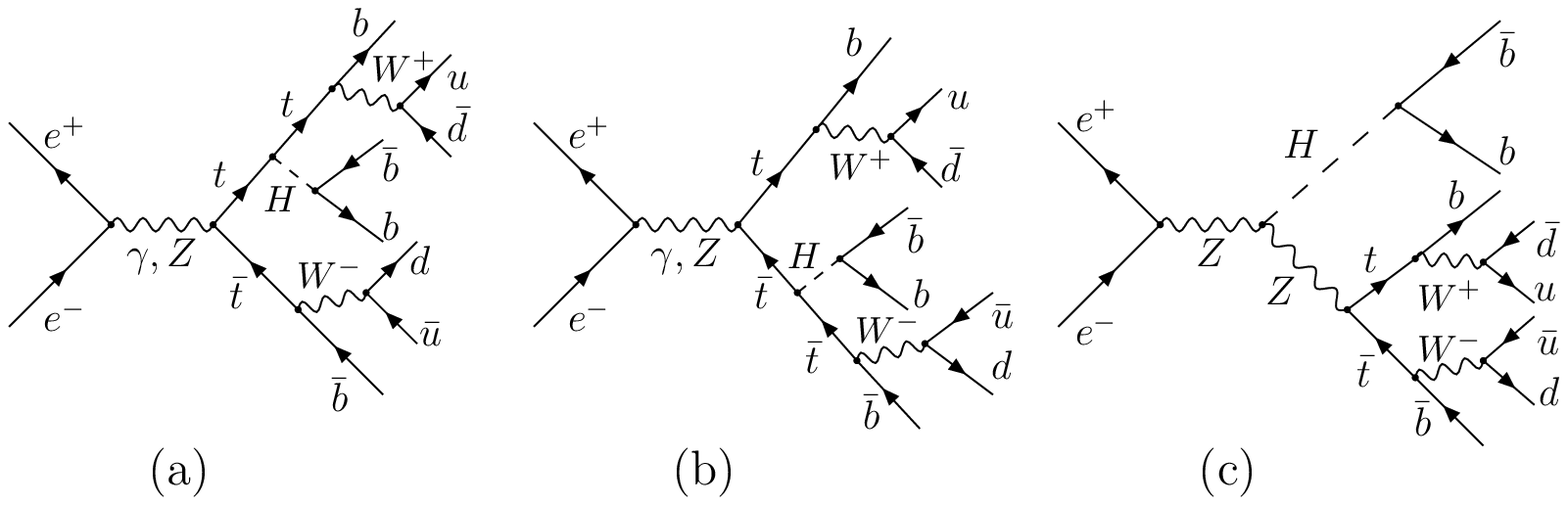}
\caption{Representative signal Feynman diagrams of reaction (\ref{uddu}) in 
the unitary gauge.
The remaining diagrams are obtained by all possible permutations of 
the two $b$ and two $\bar b$ lines. The Higgs boson coupling to electrons
has been neglected.}
\label{fig:uddu}
\end{figure*}

To be more specific, let us consider, e.g. a reaction of associated 
production of the top quark pair and Higgs boson at the ILC 
\beq
\label{eetth}
\epm \ra t \bar t H.
\eeq
Because its cross section is by far dominated by 
the Higgsstrahlung off the top quark line, reaction (\ref{eetth}) 
can be used to measure the top--Higgs Yukawa 
coupling \cite{eetth}.
As the top and antitop decay, even before they hadronize, predominantly into 
$b W^+$ and $\bar{b} W^-$, respectively, and the Higgs boson, 
dependent on its mass $m_H$, decays mostly either into a $b \bar b$-quark pair 
or an electroweak (EW) gauge boson pair
and the EW  bosons subsequently decay, each into a fermion--antifermion pair,
reaction (\ref{eetth}) will lead to reactions with either 8 or  
10 fermions in the final state. 
If $m_H < 140$~GeV, which is favoured by the direct searches at 
LEP \cite{LEPdir} and theoretical constrains in the framework 
of SM \cite{Hmass}, 
then the Higgs boson would decay mostly into a $b \bar b$-quark pair
resulting in reactions of the form
\beq
\label{ee8f}
  e^+e^-\;\; \ra \;\;  b \bar b  b \bar b f_1\bar{f'_1} f_2 \bar{f'_2},
\eeq
where $f_1, f'_2 =\nu_{e}, \nu_{\mu}, \nu_{\tau}, u, c$ and 
$f'_1, f_2 = e^-, \mu^-, \tau^-, d, s$ are the decay products  of the 
$W$-bosons coming from decays of the $t$- and $\bar t$-quark. 
Thus, in this case reaction (\ref{eetth}) 
can be detected in any of the following channels: 
the leptonic, semileptonic and  hadronic
channels, as represented by the reactions
\bea
\label{tnmn}
\epm &\ra& b \bar b b \bar b \tau^+ \nu_{\tau} \mu^- \bar \nu_{\mu},\\
\label{udmn}
\epm &\ra& b \bar b b \bar b u\bar d \mu^- \bar \nu_{\mu},\\
\label{udsc}
\epm &\ra& b \bar b b \bar b u\bar s d \bar c,
\eea
respectively. Taking into account both the EW and quantum 
chromodynamics (QCD) lowest order contributions in the unitary
gauge, with the neglect of the Yukawa couplings
of the fermions lighter than $c$ quark and $\tau$ lepton, there are 21\,214,
26\,816 and 39\,342 Feynman diagrams of reactions (\ref{tnmn}), 
(\ref{udmn}) and (\ref{udsc}), respectively. 
If both $W^+$ and $W^-$ decay into quarks of the same family we obtain,
e.g. the reaction
\bea 
\label{uddu}
\epm &\ra& b \bar b b \bar b u\bar d d \bar u
\eea
which, neglecting the light fermion masses, receives contributions 
from 185\,074 Feynman diagrams.
Most of the diagrams comprise background to 
the 20 signal diagrams representing
resonant production and decay of the top quark pair and Higgs boson. 
For illustration, the representative
signal diagrams of reaction (\ref{uddu}) are shown in Fig.~\ref{fig:uddu}.

There exists several multipurpose Monte Carlo (MC)
generators such as {\tt HELAC/PHEGAS}~\cite{HELAC/PHEGAS}, 
{\tt AMEGIC++/Sherpa}~\cite{AMEGIC++/Sherpa}, 
{\tt O'Mega/Whizard}~\cite{O'Mega/Whizard},
{\tt MadGraph/MadEvent}~\cite{MadGraph/MadEvent}, {\tt ALPGEN}~\cite{ALPGEN}, 
or {\tt CompHEP/CalcHEP}~\cite{CompHEP}.
However, one may encounter problems while trying to obtain reliable SM
predictions for multiparticle reactions as (\ref{tnmn})--(\ref{udsc}),
not to mention reaction (\ref{uddu}),
with publicly available versions of the generators.

In this lecture, the current status of {\tt carlomat}, a new program 
for automatic 
computation of the lowest order cross sections of multiparticle reactions
is described, the results of comparisons with other multipurpose 
MC programs are shown and some new results on 
reaction (\ref{uddu}) are presented.

\section{A PROGRAM}

{\tt carlomat} is a program written in Fortran 90/95.
It generates the matrix element for a user specified process and
phase space parametrizations, which are later used for the 
multichannel MC integration of the lowest order cross sections and
event generation. The program takes into account both the EW and QCD 
lowest order contributions. Particle masses are not neglected in
the program. The number of external particles is limited to 12 and
only the SM is currently implemented in the program.

{\tt carlomat} works according to the following scheme.
User specifies the process he wants to have calculated.
Then topologies for a given number of external particles are generated
and checked against Feynman rules which have been coded in the program. 
In this process,
helicity amplitudes, the colour matrix and phase space parametrizations are
generated.
Finally, they are copied to another directory where the numerical program 
can be executed.

\subsection{Generation of topologies}

\begin{figure}[h]
\vspace*{-0.3cm}
\includegraphics[scale=0.9]{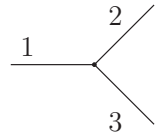}
\vspace*{-0.5cm}
\caption{The single topology of a 3 particle process.}
\label{fig:top3}
\end{figure}

Let us consider models with triple and quartic couplings.
The process of generation of topologies starts 
with 1 topology of a 3 particle process that is depicted in 
Fig.~\ref{fig:top3}.

\begin{figure*}[htb]
\vspace{3.cm}
\includegraphics{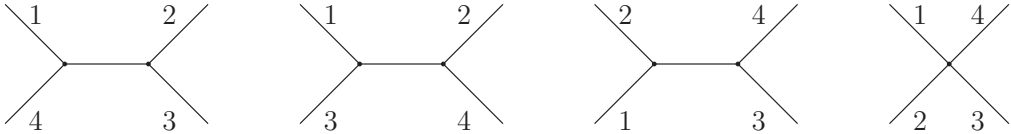}
\vspace*{-1.0cm}
\caption{The topologies of a 4 particle process.}
\label{fig:top4}
\end{figure*}

The 4 topologies of a 4 particle process which are depicted in 
Fig.~\ref{fig:top4} are obtained by attaching line No. 4 to each line 
and to the vertex of the graph in Fig.~\ref{fig:top3}.
Then the 25 topologies of
a 5 particle process are obtained by attaching line No. 5 to each line, 
including the internal ones, and to each triple vertex of the graphs 
in Fig.~\ref{fig:top4}.

The number of topologies grows dramatically with the number of external
particles.
\begin{table}[htb]
\label{table:tops}
\begin{tabular}{cr}
No. of particles & No. of topologies\\[2mm]
6  & 220 \\
7  & 2\,485 \\
8  & 34\,300\\
9  & 559\,405\\
10\,\,\,\, & 10\,525\,900\\
11\,\,\,\, & 224\,449\,225
\end{tabular}
\end{table}
For a process with $n$ external particles, topologies for $n - 1$
particles 
are generated recursively and then, while adding the $n$-th particle, 
the program checks whether a topology results in a Feynman diagram or not.
Topologies can be generated and stored on a disk prior to 
the program execution.

\subsection{Feynman diagrams}
Actual external particles are assigned to lines $1,2,3,...,n$ in a 
strict order.
Each topology is divided into two parts which are separately
checked against the Feynman rules. 
Two or three, external lines are joined by means of a triple or
quartic, vertex of the implemented model.
In this way an off shell particle, which is
represented by a spinor or polarization vector, is created.
The off shell particles and/or external particles are joined in this way
until the two parts of a considered topology are completely covered.
If they match into a propagator then the topology is accepted.
Once the topology has been accepted, the `longer' part of it is further 
divided so that the Feynman diagram is made of 3 or 4 parts, joint to form
a triple or quartic vertex of the model.
Particles defined for one Feynman diagram are used as building blocks 
of other Feynman diagrams.
The number of building blocks generated in {\tt carlomat} is usually 
smaller than in {\tt MadGraph}.

When the diagram is created, the corresponding `particles'
are used to construct the helicity amplitude, 
colour factor (matrix) and phase space parametrization which
are stored on the disk.
Once all the topologies have been checked 
subroutines for calculating the matrix element, colour matrix and phase space
integration are written.

\subsection{Helicity amplitudes}

The helicity amplitudes are computed using the routines 
developed for MC programs {\tt ee4fgamma} \cite{ee4fgamma} and
{\tt eett6f} \cite{eett6f} which have been improved and tailored to 
meet needs of the automatic generation of amplitudes.
In order to speed up the computation the MC summing over helicities has
been implemented in the program. 
However, explicit summing over helicities is also possible. While doing so,
spinors or polarization vectors representing particles, both on- and
off-shell ones, are computed only once, for all the helicities of the external
particles they are made of, and stored in arrays, which is a novel feature 
with respect to other programs, e.g. {\tt MadGraph}.

Possible poles in the propagators of unstable particles are regularized 
by constant widths which 
are introduced through the complex mass parameters
\bea
M_B^2\!&=\!&m_B^2-im_B\Gamma_B, \qquad B=W, Z, H, \nonumber \\
 \qquad M_t&=&\sqrt{m_t^2-im_t\Gamma_t},
\eea
which replace masses in the corresponding propagators
\bea
\Delta_F^{\mu\nu}(q)\!&=&\!\frac{-g^{\mu\nu}+q^{\mu}q^{\nu}/M_V^2}
                               {q^2-M_V^2},   \nonumber \\
\Delta_F(q)\!&=&\!\frac{1}{q^2-M_H^2}, \qquad
S_F(q)=\frac{/\!\!\!q+M_t}{q^2-M_t^2}, \nonumber
\eea
both in the $s$- and $t$-channel Feynman diagrams.
Propagators of a photon and gluon are taken in the Feynman gauge.
The EW mixing parameter may be defined either real, referred to as
the fixed width scheme (FWS), or complex, referred to as the complex
mass scheme (CMS)
\bea
\sin^2\theta_W=1-\frac{m_W^2}{m_Z^2} \quad {\rm or} \quad
\sin^2\theta_W=1-\frac{M_W^2}{M_Z^2}. \nonumber
\eea

The colour matrix is calculated only once at the beginning of execution 
of the numerical program after having reduced its size with the use of 
the SU(3) algebra properties.

\subsection{Phase space integration}

A dedicated phase space parametrization is generated for each Feynman
diagram.
Mappings of
the Breit-Wigner shape of the propagators of unstable particles and
$\sim 1/s$ behaviour of the photon and gluon propagators
are performed.
Dedicated treatment of soft and collinear external photons,
as well as $t$-channel photon/gluon exchange is envisaged.

The phase space parametrizations are incorporated into a multichannel
MC integration routine for calculating total and differential cross sections.
The integration is performed iteratively. It
starts with equal weights for all the kinematical channels and 
the weights with which each kinematical channel contributes to the integral 
are determined anew after every iteration.
{\tt carlomat} can be used as MC generator of unweighted events as well.

\section{TESTS}

Matrix elements of many reactions with 6 particles and several reactions 
with 7 particles in the final state have been checked against 
{\tt MadGraph} for randomly selected sets of momenta. An agreement 
better than 13 digits has been found.

Total cross sections of reactions $\epm \ra$ 4 fermions and $\epm \ra$ 
4 fermions and a photon have been checked against {\tt ee4fgamma}
\cite{ee4fgamma}, and of $\epm \ra$ 6 fermions, relevant for 
the top quark pair production and decay, have been compared with {\tt eett6f}.
The results have agreed within one standard deviation of the MC
integration. 

Moreover, checks against other MC programs have been made. 
Cross sections of the following top quark 
pair production reactions at the ILC:
\bea
\label{uudd}
\epm &\ra& b\bar{b} u \bar{d} d\bar{u} 
,\\
\label{uden}
\epm &\ra& b\bar{b} u\bar{d} e^-\bar{\nu}_e,\\
\label{enmn}
\epm &\ra& b\bar{b} \nu_e e^+ \mu^-\bar{\nu}_{\mu},\\
\label{mnmn}
\epm &\ra& b\bar{b} \nu_{\mu}\mu^+ \mu^-\bar{\nu}_{\mu}
\eea
computed with {\tt carlomat} are compared against of the corresponding
results of {\tt HELAC/PHEGAS} and
{\tt AMEGIC++} \cite{Gleisberg} in Table~\ref{table:comps}. 
The cuts and initial parameters are those of \cite{Gleisberg} and,
as in \cite{Gleisberg},
the cross sections have been calculated with $10^6$ calls 
in the MC integration, before the cuts have been applied.
\begin{table}[htb]
\caption{Cross sections in fb of (\ref{uudd})--(\ref{mnmn}) at $\sqrt{s}=
360$~GeV (first row) and $\sqrt{s}=500$~GeV (second row). The cuts and
initial parameters are as those of \cite{Gleisberg}. The numbers in parentheses
show the uncertainty of the last decimals.}
\label{table:comps}
\begin{tabular}{ccccc}
\hline \\[-1.5mm]
Reac. & {\tt carlomat} 
& {\tt AMAGIC++} & {\tt HELAC} \\[1.5mm]
\hline \\[-1.5mm]
(\ref{uudd})
& 32.98(11) & 32.90(15) & 33.05(14) \\
& 50.31(19) & 49.74(21) & 50.20(13) \\[1.5mm]
(\ref{uden})
& 11.448(26) & 11.460(36) & 11.488(15) \\
& 17.424(56) & 17.486(66) &  17.492(41) \\[1.5mm]
(\ref{enmn})
& 3.843(5) & 3.847(15) & 3.848(7) \\
& 5.856(11) & 5.865(24) & 5.868(10) \\[1.5mm]
(\ref{mnmn})
& 3.837(5) & 3.808(16) & 3.861(19) \\
& 5.834(10) & 5.840(30) & 5.839(12) \\[1.5mm]
\end{tabular} 
\end{table}
Satisfactory agreement also for all the other cross sections 
presented in \cite{Gleisberg} for about 80 
reactions has been found. However, for some
reactions containing gluons, or $\epm$ pair in the final state, more calls
had to be used in {\tt carlomat} in order to match the corresponding 
precision of \cite{Gleisberg}, as the appropriate mappings have not yet been
implemented in the program.
Cross sections of reaction (\ref{udmn}) without the gluon exchange 
contributions computed with {\tt carlomat}
and {\tt Whizard} are compared 
in Table~\ref{table:whiz}. Again a satisfactory agreement can be seen.
\begin{table}[htb]
\caption{Cross sections in ab of (\ref{udmn}) without the gluon exchange
contributions calculated with {\tt carlomat} and {\tt Whizard}.
The cuts and initial parameters are as those of \cite{KS}. 
The numbers in parentheses
show the uncertainty of the last decimal.}
\label{table:whiz}
\begin{tabular}{ccc}
\hline\\[-1.5mm]
$\sqrt{s}$ [GeV]
& {\tt carlomat} & {\tt Whizard} \\[1.5mm]
\hline\\[-1.5mm] 
500 & 7.80(3) & 7.76(2) \\
800 & 66.8(1) & 67.3(1) \\
1000\,\,\, & 61.4(1) & 61.8(2) \\
2000\,\,\, & 28.5(1) & 28.1(3) \\
\end{tabular} 
\end{table}
\section{SAMPLE RESULTS}
The capability of {\tt carlomat} to handle multiparticle reactions with 
large numbers of the Feynman diagrams was demonstrated
in \cite{KS}, where the off resonance background effects in
the associated top quark pair and Higgs boson production at the linear
$\epm$ collider were studied. In \cite{KS}, the effects were shown
in reactions (\ref{tnmn})--(\ref{udsc}). Here, we will address
the question of the off resonance background
contributions to the associated top quark pair and Higgs boson production 
in reaction (\ref{uddu}) which is the hadronic detection channel 
of (\ref{eetth}) that, in the unitary gauge and neglecting the Yukawa 
couplings to light fermions, 
receives contributions from 185\,074 lowest order 
Feynman diagrams.

As in \cite{KS}, let us identify jets with their original partons 
and define the following 
cuts on an angle between a quark and a beam and an angle between two quarks
\beq
\label{angcuts}
5^{\circ} < \theta (q,\mathrm{beam}) < 175^{\circ}, \qquad
\theta (q,q') > 10^{\circ}
\eeq
and a cut on a quark energy
\beq
\label{encut}
E_{q} > 15\;{\rm GeV},
\eeq
which should allow to detect events with 8 separate jets. Moreover, in
order to reconstruct $W$ bosons, top quarks and the Higgs boson let us 
assume 100\% efficiency of $b$ tagging and 
impose the following invariant mass cuts:\\
a cut on the invariant mass of two non $b$ jets
\beq
\label{cutmw}
60\;{\rm GeV} < \left[\left(p_{\sim b_1}+p_{\sim b_2}\right)^2\right]^{1/2} < 90
\;{\rm GeV},
\eeq
a cut on the invariant mass of a $b$ jet, $b_1$, and two non 
$b$ jets, $b_{\sim b_1}, b_{\sim b_2}$, 
\beq
\label{cutmt}
\left|\left[\left(p_{b_1}+p_{\sim b_1}+p_{\sim b_2}\right)^2\right]^{1/2} 
- m_t\right| < 30\;{\rm GeV}
\eeq
and an invariance mass cut on two $b$ jets, $b_3$ and $b_4$, 
\beq
\label{cutmh}
\left|\left[\left(p_{b_3}+p_{b_4}\right)^2\right]^{1/2} - m_h\right|
< m_{bb}^{\rm cut}, 
\eeq
with $m_{bb}^{\rm cut} = 20$~GeV, 5~GeV, 1~GeV.

The total cross section of reaction (\ref{uddu}) calculated
with the complete set of the lowest order Feynmamn diagrams,
$\sigma_{\rm all}$, and with the 20 signal diagrams, representatives of
which are depicted in Fig.~\ref{fig:uddu},
$\sigma_{\rm sig.}$, with cuts (\ref{angcuts})--(\ref{cutmh})
are shown in Table~\ref{table:uddu}. The initial parameters used are 
those of \cite{KS}. 
\begin{table}[htb]
\caption{The cross sections $\sigma_{\rm all}$ and $\sigma_{\rm sig.}$
of (\ref{uddu}) for 3 different values of the invariant mass cut
$m_{bb}^{\rm cut}$ of (\ref{cutmh}). The other cuts are given by 
(\ref{angcuts})--(\ref{cutmt}) and initial parameters used are as those of 
\cite{KS}. The numbers in parentheses show the uncertainty of the last 
decimal.}
\label{table:uddu}
\begin{tabular}{ccrr}
\hline\\[-1.5mm]
$\sqrt{s}$ [GeV] & $m_{bb}^{\rm cut}$ [GeV]
& $\sigma_{\rm all}$ [ab] & $\sigma_{\rm sig.}$ [ab] \\[1.5mm]
\hline\\[-1.5mm] 
500 & 20\,\, & 13.46(5) & 8.72(1) \\
    & 5 & 10.12(4) & 8.70(1) \\
    & 1 &  8.98(4) & 8.67(1) \\[1.5mm]
800 & 20\,\, & 164.5(4) & 128.6(1) \\
    & 5 & 139.5(4) & 128.2(1) \\
    & 1 & 129.6(2) & 127.7(1) \\[1.5mm]
1000\,\,\,\,& 20\,\, & 137.9(3) & 109.2(1) \\
    & 5 & 117.7(5) & 109.1(1) \\
    & 1 & 110.6(5) & 108.6(1) \\[1.5mm]
2000\,\,\,\,& 20\,\, & 44.2(2) & 36.17(4) \\
    & 5 & 38.2(2) & 36.27(4) \\
    & 1 & 36.5(1) & 36.14(4) 
\end{tabular} 
\end{table}
We see that invariant mass cut (\ref{cutmh}), which has been imposed in
order to identify
the $b\bar b$ quark pair coming from the Higgs boson decay, reduces
the off resonance background very efficiently while it practically does
not alter the signal cross section.

\section{CONCLUSIONS AND OUTLOOK}

{\tt carlomat} can be used for automatic computation of 
cross sections of multiparticle reactions as it has been demonstrated
in \cite{KS} and in Section~4. It can be used as an MC generator 
of unweighted events, too.

In spite of all the successful checks presented in Section~3, further 
thorough comparisons with other existing MC generators 
still should be done.

Interfaces to PDF's, or ISR within the structure function
approach are practically ready.
Interfaces to parton shower and hadronization programs should 
be worked on.
Extensions of SM can be implemented and the corresponding lowest order
cross sections can be calculated in a fully automatic way.

Leading SM radiative corrections can be implemented, if
corresponding subroutines are provided as it was done, e.g. in
\cite{radcor}, where factorizable EW $\cal{O}(\alpha)$ corrections
were included for reactions $e^+e^- \ra$ 6 fermions relevant for the top 
quark pair production and decay at a linear $e^+e^-$ collider.

\end{document}